\documentclass[preprint2,trackchanges]{aastex62}

\usepackage{epstopdf}
\received{March 1, 2019}
\revised{December 9, 2019}
\accepted{December 27, 2019}
\published{February 10, 2020}
\submitjournal{ApJ}

\shorttitle{Mode Switching of PSR J0614+2229}
\shortauthors{Y. R. Zhang et al.}

\turnoffeditone
\begin{document}

\title{Multifrequency Study on the Mode Switching of PSR J0614+2229}

\correspondingauthor{H. G. Wang}
\email{hgwang@gzhu.edu.cn}

\author[0000-0002-8881-4126]{Y. R. Zhang}
\affiliation{School of Physics and Electronic Engineering, \\
Guangzhou University, \\
510006 Guangzhou, China}
\affiliation{Department of Physics and Astronomy, \\
University of Padova, \\
Via F. Marzolo 8, I-35131 Padova, Italy}

\author[0000-0002-2044-5184]{H. G. Wang}
\affil{School of Physics and Electronic Engineering, \\
Guangzhou University, \\
510006 Guangzhou, China}
\affil{Xinjiang Astronomical Observatory, \\
Chinese Academy of Sciences, \\ 
150 Science 1-Street, Urumqi, Xinjiang 830011, China}

\author{X. J. Huang}
\affil{College of Physical Science and Technology, \\
Central China Normal University, \\
430079, Wuhan, Hubei, China}

\author{J. L. Chen}
\affil{Department of Physics and Electronic Engineering, \\
Yuncheng University, \\
044000, Yuncheng, Shanxi, China}
\nocollaboration



\begin{abstract}

  The mode switching phenomenon of PSR J0614+2229 was studied by using the archived observations at 686, 1369 and 3100 MHz with the Parkes radio telescope which have not been published before, and combining existing observations from the literature. Over a wide frequency range from 327 to 3100 MHz, the pulsar switches between one mode occurring earlier in pulse phase (mode A) and the other mode appearing later in phase (mode B), with a generally stable phase offset between their profile peaks. The two modes are found to be different in the following aspects. (1) Mode A has a flatter spectrum than mode B, with a difference in the spectral index of about 0.5. This accounts for the phenomenon that the flux ratio between the modes A and B increases with frequency, and mode A becomes stronger than mode B above $\sim500$ MHz. (2) For mode B, the flux density of the subintegrated profile is anticorrelated with the emission phase, indicating that the emission from earlier phases is relatively stronger than that from later phases; such an anticorrelation is not observed in mode A. (3) The frequency dependence of the FWHM of the two modes are opposite each other; namely, the FWHM of mode A increases with frequency, while that of mode B decreases with frequency. A possible interpretation is suggested that the longitudinal spectral variation across the two beams may be opposite each other.

\end{abstract}



\defcitealias{1980ApJ...239..310F}{FB80}
\defcitealias{2014MNRAS.439.3951S}{SLR14}
\defcitealias{2016MNRAS.462.2518R}{RSL+16}
\defcitealias{2018MNRAS.473.4436J}{JSK+18}
\defcitealias{2017MNRAS.470.2659G}{GKK+17}

\section{Introduction} \label{sec:Intro}
PSR J0614+2229 (B0611+22), with a period of 0.33 s and a characteristic age of 90,000 yr, is a young pulsar discovered by \citet{1972Natur.240..229D}. Observations in 1970s revealed that it had high timing noise \citep{1980ApJ...237..206H}. Its profile variability at 430 MHz was studied by  Ferguson \& Boriakoff (\citeyear{1980ApJ...239..310F}, hereafter \citetalias{1980ApJ...239..310F}), who found that the pulse peaks of short integrations (in minutes) vary in phase by as much as $\sim$2 milliseconds ($\sim$2\degr), and suggested that a brighter pulse might come at a later phase. Those authors suspected that the profile variation might be caused by mode switching. This conjecture was confirmed by a later study at the same frequency \citep{nowakowski_1992}. It was found that the pulsar switches between two emission modes, of which the average profiles are of Gaussian shape but with different width, brightness, and pulse phase. In particular, the mode \edit1{that peaks} at a later phase (hereafter mode B) appeared to be stronger than the other mode \edit1{that peaked} at an earlier phase (hereafter mode A).

Recent studies at multiple frequencies have revealed new features of the mode switching of this pulsar. Seymour et al. (\citeyear{2014MNRAS.439.3951S}, hereafter \citetalias{2014MNRAS.439.3951S}) found that while the emission of mode A is steady, the intensity of mode B is largely enhanced at 327 MHz, and appears to systematically decay from pulse to pulse. Mode B is regarded as a bursting mode at 327 MHz, but at 1400 MHz the two modes show comparable intensities. Besides the brightness, the pulse width and phase also vary with frequency. Unlike \citetalias{1980ApJ...239..310F}, \citetalias{2014MNRAS.439.3951S} did not find any subpulse drifting in either the modes. Using the Low-Frequency Array (LOFAR), Arecibo and Green Bank Telescopes, Rajwade et al. (\citeyear{2016MNRAS.462.2518R}, hereafter \citetalias{2016MNRAS.462.2518R}) observed this pulsar for a much longer time, revealing that the bursts have a rough periodicity of $\sim$2500 pulse periods ($\sim$14 minutes). Simultaneous observations at 150/327 MHz and 150/820 MHz \edit1{showed} that the bursts are correlated at 327 and 150 MHz while anticorrelated at 820 and 150 MHz; namely, mode B is stronger (enhanced by the bursting emission) at both 150 and 327 MHz, while mode A becomes stronger at 820 MHz. \citetalias{2016MNRAS.462.2518R} suggested that this might be caused by different spectral indices of the two modes.

From the above results, there seems to be an intriguing frequency dependence of the relative brightness between the two modes. First, mode A is much weaker than mode B at \edit1{150-327} MHz; then the ratio between modes A and B keeps increasing, leading mode A to {surpass} mode B at 820 MHz. However, this tendency is reversed, and mode B {becomes} comparable with mode A at 1.4 GHz. This unusual behavior requires further investigation at higher frequencies.

In this paper, by using the archived data observed with the Parkes radio telescope in the 10, 20, and 50 cm bands and collecting the relevant results at other frequencies from the literature, we focus on the frequency dependence of pulse width, brightness and phase offset between the two modes over a decade of frequency range from $\sim$0.3 to $\sim$3 GHz. Among the three bands, the 10 and 50 cm bands have not been used to publish the mode switching of this pulsar before. Details of observations are given in Section \ref{sect:Obs}. Results are presented in Section \ref{sect:Rslt}. The implication of our result to the emission beam is discussed in Section \ref{sect:Impl}. Conclusions and discussions are presented in Section \ref{sect:Disc}.

\section{Observations} \label{sect:Obs}

The archived data of this pulsar since 2005 are available from the Parkes pulsar data archive \citep{2011PASA...28..202H}. Although this pulsar has been observed with the Parkes radio telescope many times, most observations are for the purpose of timing and only last for a few minutes, which can barely record a mode switching. Fortunately, we found a 50-min simultaneous observation in both the 50 and 10 cm bands with a dual frequency receiver system and an 18-min observation in the 20 cm band with the H-OH receiver, which can be used to analyze the frequency dependence of the moding behavior of this pulsar. These two observations were carried out on 2005 October 15 and 24, respectively. The data were recorded with a wide-band correlator (WBCORR) at all the three frequencies. Table \ref{tab:Obs} summarizes the ID of the observations used in this paper, the MJD, the receiver, the observational frequency, the band width ($\Delta\nu$), the band width of each channel ($\Delta\nu_{\rm ch}$), the integration time of each subintegration ($T_{\rm sub}$), the number of subintegrations ($N_{\rm sub}$), and the number of bins in each profile ($N_{\rm bin}$).

\begin{deluxetable*}{ccccccccc}[htbp]
\tablecaption{Summary of Parkes observations \label{tab:Obs}}
\tablecolumns{9}
\tablenum{1}
\tablewidth{0pt}
\tablehead{
\colhead{ID}&
\colhead{MJD}&
\colhead{Receiver} & \colhead{$\nu$} & \colhead{$\Delta\nu$} & \colhead{$\Delta\nu_{\rm ch}$} & \colhead{$T_{\rm sub}$} & \colhead{$N_{\rm sub}$} & \colhead{$N_{\rm bin}$} \\
\colhead{} & \colhead{(d)} & \colhead{} & \colhead{(MHz)}  & \colhead{(MHz)} & \colhead{(MHz)} & \colhead{($s$)} & \colhead{} & \colhead{}
}
\startdata
2005a & 53658.80926 &5010CM &686 &256 &0.125 &59.6209 &51&1024 \\
2005b & 53667.80810 &H-OH   &1369&256 &0.125 &59.9563 &18&1024 \\
2005c & 53658.80926 &1050CM &3100&1024&1.0   &59.6209 &51&1024 \\
\enddata
\end{deluxetable*}

All of the data were reduced with the PSRCHIVE pulsar data analysis system\footnote{\url{http://psrchive.sourceforge.net/}} \citep{2004PASA...21..302H}. The radio frequency interference (RFI) was removed with the PSRCHIVE program \emph{paz} by median filtering and the edge channels with little signal were zero weighted in the frequency domain. The removal of the RFI contaminating a considerable fraction of the lower frequency channels results in a weighted frequency of 686 MHz in the 50 cm band, while for the 10 and 20 cm bands with little RFI, the central frequencies of 3100 and 1368 MHz are taken. The flux densities were calibrated from the radio source Hydra A \citep{2012AR&T....9..237V}. The reduced datasets are then used in the following analysis\footnote{The software Python (\url{https://www.python.org/}) with packages NumPy (\url{http://www.numpy.org/}), pandas (\url{http://pandas.pydata.org/}), SciPy (\url{https://www.scipy.org/}), Matplotlib (\url{https://matplotlib.org/}) and Uncertainties (\url{http://pythonhosted.org/uncertainties/}) were used in the analysis.}.

\section{RESULTS} \label{sect:Rslt}

\begin{figure*}[htbp]
\plotone{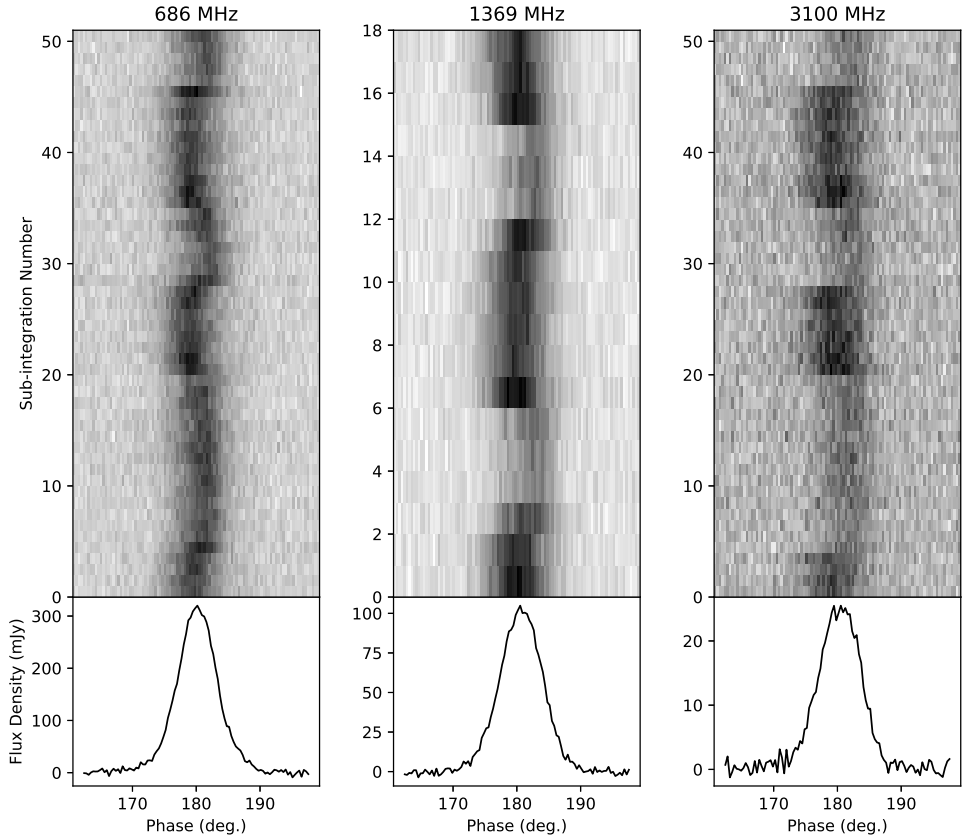}
\caption{Grayscale plots of the subintegration sequences and the mean-pulse profiles for the Parkes observations. \edit1{Each subintegration is 1 min.} The sequences at 686 and 3100 MHz were observed simultaneously. The mean pulse profiles were manually aligned by placing the peaks at the phase 180\degr. \label{fig:raw}}
\end{figure*}

The grayscale plots of the subintegration profiles at three frequencies are shown in Fig. \ref{fig:raw}. Modes A and B can be easily identified according to their phase offset. From the panels of 686 and 3100 MHz, one can see that the mode switching occurs simultaneously at these two frequencies. It is noticed that mode A is brighter than mode B at all the three frequencies, \edit1{and the difference becomes more obvious at higher frequencies}. This is consistent with the observation of \citetalias{2016MNRAS.462.2518R} at 820 MHz, but contrary to \edit1{all other multiband observations in which the burst pulses usually appear in B mode (see \citetalias{1980ApJ...239..310F}; \citealt{nowakowski_1992}; \citetalias{2016MNRAS.462.2518R}; \citetalias{2014MNRAS.439.3951S} for details)}. It seems that mode A has a flatter spectrum than mode B. Below, we perform a quantitative analysis on the spectra of the two modes. Besides, Figs. 5 and 6 in \citetalias{2014MNRAS.439.3951S} show that the pulse width of mode B decreases with frequency, and the stronger pulses of mode B tend to appear at earlier phases at both 327 MHz and 1.4~GHz. In the following, these tendencies are also studied quantitatively.

\subsection{Spectral difference between the two modes}
\label{subsec:Spec}

\begin{figure*}[htbp]
  \plotone{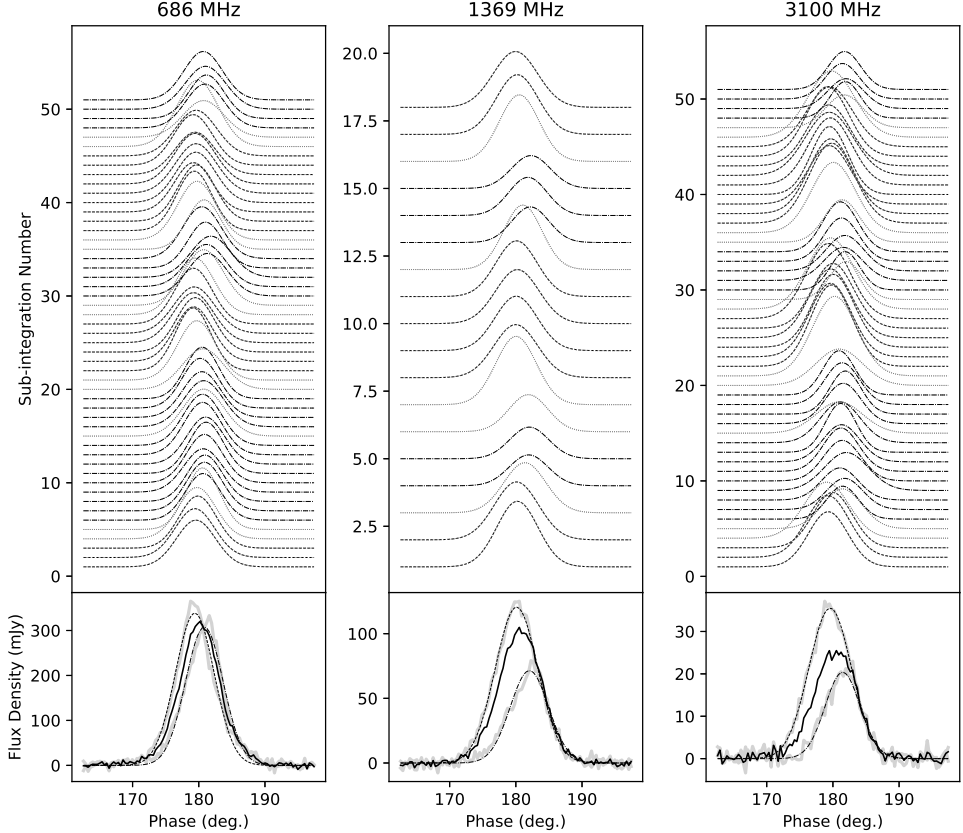}
   \caption{Best-fit Gaussians for the sub-integrated pulse profiles at three frequencies. The peak intensities of the Gaussian profiles are multiplied by the same factor at each frequency. The dashed and dashed-dotted curves stand for the profiles of modes A (earlier in peak phase) and B (later in peak phase), while the gray dotted curves are the removed profiles. The bottom panels show both the original mean-pulse profiles (gray solid lines) and the best-fit Gaussians (dashed and dashed-dotted lines) for modes A and B. The average pulse profiles of total emission are also shown by the solid black lines. \label{fig:fit}}
\end{figure*}

 To obtain the integrated profiles of the two modes for each dataset, firstly we divided the subintegrations into two groups according to their phases. Most subintegrations can be easily identified, except for those during which the switching occurred so that they consist of pulses of the two modes. Therefore, this kind of subintegrations were removed. As pointed out by other authors, the integrated pulse profile is well described by a Gaussian (\citealt{1999ApJS..121..171W}; \citetalias{2014MNRAS.439.3951S}). We also found this for both of the modes at all the three frequencies, except that at 686 MHz the profiles deviate slightly from the Gaussian shape near the profile peaks of the two modes and at the trailing edge of the mode A profile. The best-fit Gaussians are presented in Fig. \ref{fig:fit}. The subintegrations were separated into mode A and mode B, and then superposed to form the mean pulse profiles for both of the modes. The obtained profiles and the best-fit Gaussians, together with the mean-pulse profiles of total emission, are shown at the bottom of each panel in Fig. \ref{fig:fit}.

\begin{figure}[htbp]
  \plotone{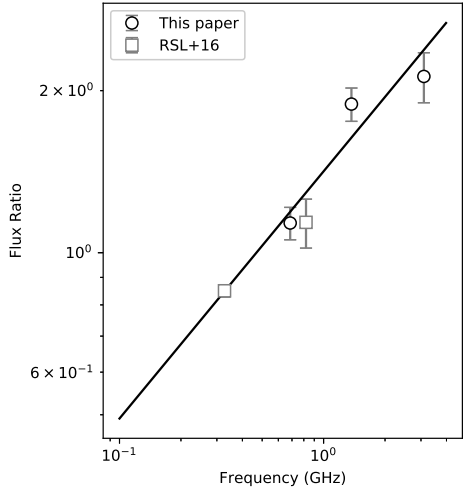}
   \caption{Frequency dependence of the flux ratio $I_{\rm A}/I_{\rm B}$ between modes A and B. \label{fig:flx}}
\end{figure}

The flux densities were measured \edit1{with the PSRCHIVE programs \emph{paas} and \emph{psrflux}}, and are listed in Table \ref{tab:Flx}, where $I_{\rm T}$, $I_{\rm A}$ and $I_{\rm B}$ represent the total flux, the fluxes of modes A and B, respectively, and $I_{\rm A}/I_{\rm B}$ is the flux ratio between the two modes. Fluxes at \edit1{327 and 820 MHz} were also measured for both of modes by using the published profiles in \citetalias{2016MNRAS.462.2518R}. 
It is obvious that the flux ratio $I_{\rm A}/I_{\rm B}$ is an increasing function of the frequency, with a best-fit \edit1{power-law relationship to be $I_{\rm A}/I_{\rm B}=1.42(5)\nu^{0.47(5)}$}, as shown by Fig.~\ref{fig:flx}. This strongly suggests that mode A has a flatter spectrum than mode B.

\begin{deluxetable*}{llcrr rrrl}[htbp]
\tablecaption{Flux densities of modes A and B in different observations \label{tab:Flx}}
\tablecolumns{9}
\tablenum{2}
\tablewidth{0pt}
\tablehead{
\colhead{UT}&
\colhead{MJD}&
\colhead{Telescope} & \colhead{$\nu$} & \colhead{$I_{\rm T}$} & \colhead{$I_{\rm A}$} & \colhead{$I_{\rm B}$} & \colhead{$I_{\rm A}/I_{\rm B}$} &\colhead{Ref.} \\
\colhead{}&
\colhead{(d)} & \colhead{} & \colhead{(MHz)}  & \colhead{(mJy)} & \colhead{(mJy)} & \colhead{(mJy)} & \colhead{} & \colhead{}
}
\startdata
 2005 Oct 15 & 53658.80926 & Parkes    &  686    &{6.59(2)}	 &{6.85(5)}	   &{6.03(4)}     & {1.14(8)} & 2005a \\
 2005 Oct 24 & 53667.80810 & Parkes    & 1369    &{2.30(1)}	 &{2.68(2)}	   &{1.42(2)}     & {1.88(13)} & 2005b \\
 2005 Oct 15 & 53658.80926 & Parkes    & 3100    &{0.564(5)}    &{0.785(8)}	   &{0.369(6)}    & {2.13(23)} & 2005c \\
 2014 Apr 9  & 56756       & LOFAR          & 150     &88(44)    &  --     &  --       & --  & \citetalias{2016MNRAS.462.2518R}  \\
 2014 Apr 9  & 56756       & Arecibo        & 327     &13.8(8)   &11.1(1)   &13.1(3)     &0.85(2) & \citetalias{2016MNRAS.462.2518R} \\
 2014 Apr 9  & 56756       & Green Bank     & 820     &16.1(1.7) &18.7(1.9)   &16.4(2)     &1.14(12) & \citetalias{2016MNRAS.462.2518R} \\
 2016 Jan 4  & 57391   & LOFAR    & 150     &75(32)      &  --        &  --        &  -- & GKK+17  \\
 2007-2016  &  --     & Parkes   & 728     &8.8(8)      &  --        &  --        &  -- & JSK+18  \\
 2007-2016  & --      & Parkes   & 1382    &3.3(2)      &  --        &  --        &  -- & JSK+18  \\
 2007-2016  & --      & Parkes   & 3100    &0.76(6)     &  --        &  --        &  -- & JSK+18  \\
\enddata
\tablecomments{The values in brackets stand for the errors in the last digits.
{The fluxes  of this work were obtained with the PSRCHIVE program \emph{paas} and \emph{psrflux}.} The LOFAR data observed in 2016 is the weighted mean of two fitted flux densities of the Cycle 5 observation given in the paper. As for the others, the data were adopted directly from the literature or estimated by {fitting Gaussian function to} the flux-calibrated profiles published in the literature. References: RSL+16 \citep{2016MNRAS.462.2518R}, GKK+17 \citep{2017MNRAS.470.2659G}, and JSK+18 \citep{2018MNRAS.473.4436J}.}

\end{deluxetable*}

\begin{figure*}[htbp]
  \plotone{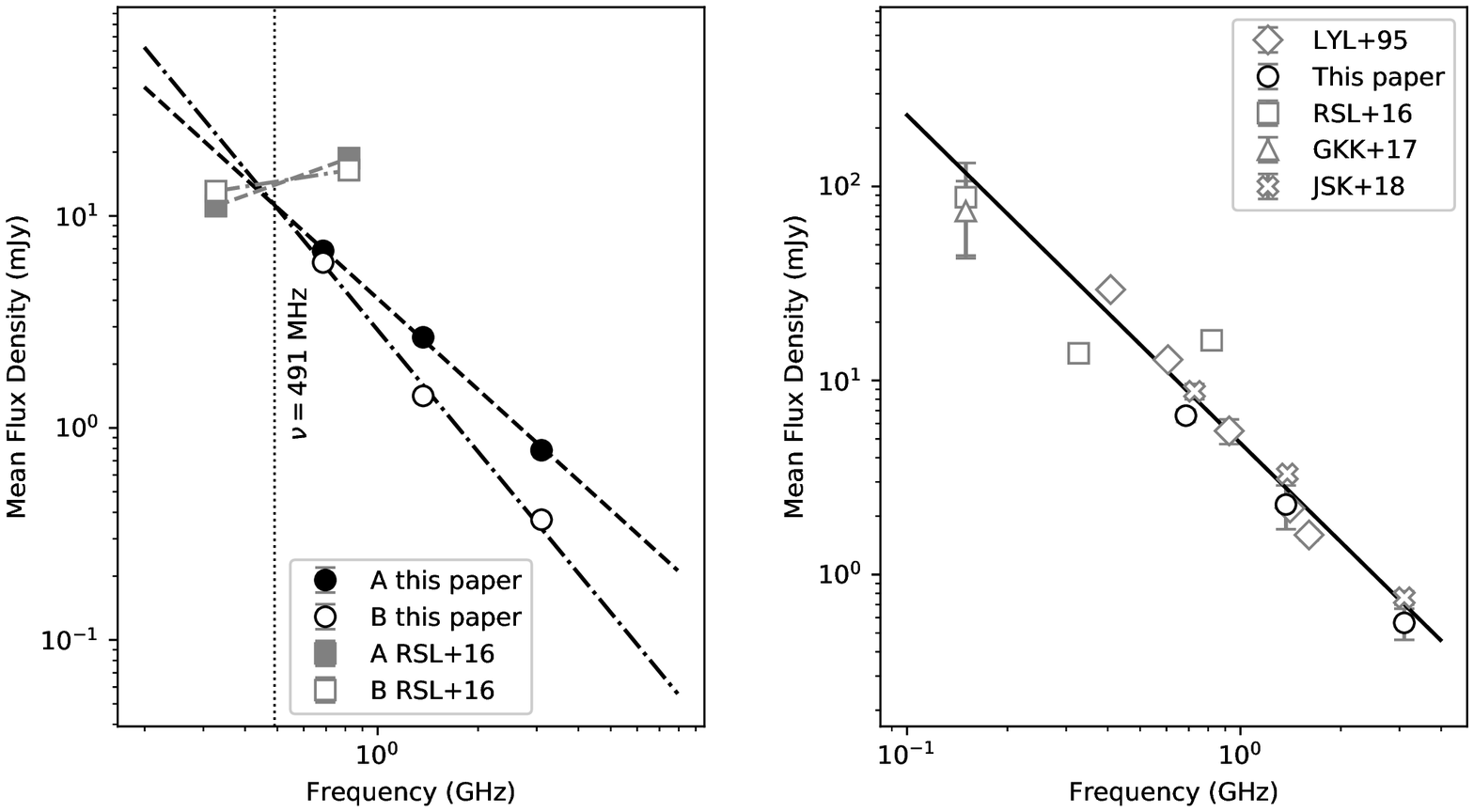}
   \caption{Left: the spectra of modes A and B for two groups of observations. The dash and dash-dotted black lines show the best-fit power laws for our data. The vertical line represents the transition frequency above which mode A becomes stronger than mode B. Although power-law fitting can not be performed to the data of \citetalias{2016MNRAS.462.2518R}, the trends of the two modes (gray lines) show similar spectral difference and transition frequency. Right: the spectrum of total emission by combining our data with those from literature. The line is the best-fit power law.
   \label{fig:spc}}
\end{figure*}

\edit1{The difference can be seen directly from the spectra of the two modes by using their intensities at 686, 1369, and 3100 MHz, as shown in Fig. \ref{fig:spc}. Obviously, mode A (solid symbols) has a flatter spectrum than mode B (open symbols). Employing power-law fitting to the spectra, the difference in the spectral indices of the two modes is measured as 0.48(1), which is well consistent with the above result of $0.47(5)$. The two power laws intersect at $\sim500$ MHz, which could be regarded as a transition frequency through which mode A becomes stronger than mode B.} This is consistent with the observations that mode A was weaker than mode B at 430 MHz (\citetalias{1980ApJ...239..310F}; \citealt{nowakowski_1992}).

\edit1{For the datasets of \citetalias{2016MNRAS.462.2518R}, $I_{\rm A}$ and $I_{\rm B}$ were only available at two frequencies (327 and 820 MHz). The number of data points is not enough to perform a least-square fitting, but they can be used to check the tendency. Comparing with the Parkes data set, in spite of the opposite sign of the slope rate, the data here do show two similarities, i.e. a flatter spectrum for mode A and a transition frequency close to $\sim500$ MHz. As for the positive sign, it could be attributed to a turnover spectrum suggested by \citetalias{2016MNRAS.462.2518R}, who used four flux measurements between 300 MHz and 1.4~GHz by combining with their own observations at 327 MHz and 820 MHz and the historical published data at the other two frequencies. We have collected a number of flux data ($I_{\rm T}$) over a much wider frequency range from the literature, which are listed in Table \ref{tab:Flx} and plotted in the right panel of Fig. \ref{fig:spc}. The spectrum simply follows a power law with the best-fit spectral index of -1.69$\pm$0.03, and the turnover is no longer seen when using a large sample of flux measurements. }

\edit1{A factor that needs to be checked is whether refractive and diffractive interstellar scintillation may affect the flux and spectral estimates. It is known that refractive scintillation \citep{1994MNRAS.269.1035G} has a timescale of the order of days and diffractive scintillation \citep{2011A&A...534A..66L} can be of the order of minutes. \citetalias{2016MNRAS.462.2518R} has excluded the effect of both kinds of scintillation on their flux estimates by comparing with the scintillation timescales with the total integration time. In this paper, the observation at 1369 MHz was carried out nine days later than those at 686/3100 MHz, which requires a check for refractive scintillation. In Table 2, the flux estimates with 1$\sigma$ errors by \citetalias{2018MNRAS.473.4436J} were obtained via long-term statistics at 728, 1382 and 3100 MHz. Our flux estimates are generally consistent with those of \citetalias{2018MNRAS.473.4436J}, indicating that refractive scintillation may not dramatically modulate the flux. As to the diffractive scintillation, because this pulsar has a large DM ($\sim96$~pc cm$^{-3}$), the estimated scintillation bandwidths are 2 kHz, 46 kHz, and 1.5 MHz for the observing frequencies 686, 1400, and 3100 MHz, respectively, by using the equations in Section 2.3 in \citet{2017MNRAS.472.1458D}. These scintillation bandwidths are much smaller than the observed bandwidths. Hence, we could neglect the effect of diffractive scintillation.}

\subsection{Frequency dependence of phase offset and pulse width}
\label{subsec:Phase}

\begin{figure*}[htbp]
  \plotone{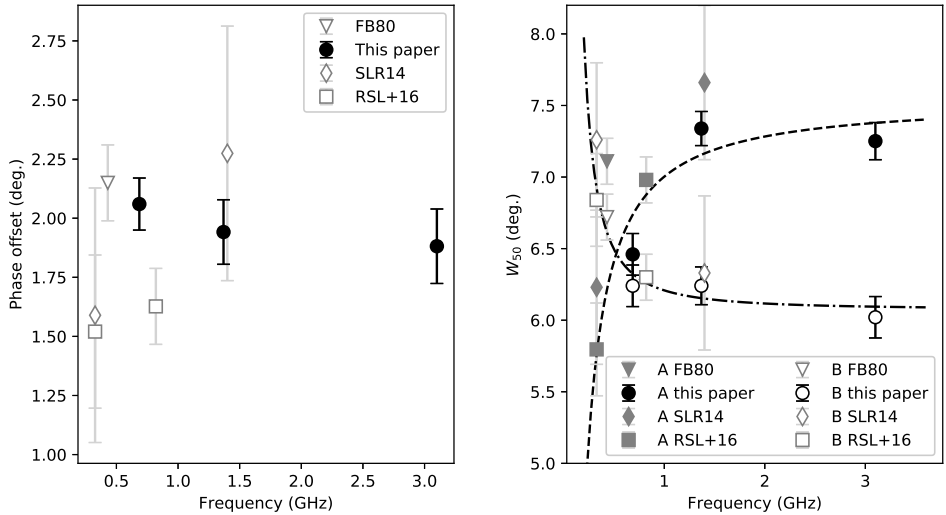}
   \caption{Left: phase offset between the pulse peaks of the two modes vs. frequency. The value of the black dot at 686 MHz is the weighted mean of $\Delta\phi$ and $\Delta\mu$ in Table 3. Right: FWHM vs. frequency for modes A (solid symbols) and B (open symbols). The best-fit Thorsett relationships were plotted as the dashed and dashed-dotted curves. The data points at 430MHz (triangles)  are plotted for reference, but not used in the fit, because the published profiles might not be well separated in modes. See the text for details. \label{fig:wph}}
\end{figure*}

Using the best-fit Gaussians for the mean-pulse profiles of the two modes, we measured the FWHM ($W_{50}$) and the phase offset between the two peaks ($\Delta\mu$). Unlike the profiles at 1369 and 3100 MHz, the profiles of the two modes at 686 MHz slightly deviate from the Gaussian shape around the peak. Therefore, the peak offset ($\Delta\Phi$) is also measured by using the original profiles at this frequency, which is larger than the value measured with the best-fit Gaussian profiles, but with a large uncertainty due to low signal-to-noise ratio. We also measured $\Delta\Phi$ and $W_{50}$ for the original profiles (not best-fit Gaussians) at 327 and 1400 MHz published by \citetalias{2014MNRAS.439.3951S} and profiles at 327 and 820 MHz published by \citetalias{2016MNRAS.462.2518R} ($\Delta\Phi$). \citetalias{1980ApJ...239..310F} did not publish profiles for separated modes at 430 MHz, but they gave the peak offsets for a couple of profiles integrated over 4.5 minutes each, among which the largest is 2\fdg15. This offset could be regarded as a good approximation for the real one,  \edit1{because in most of the subintegrations the emission should be entirely in mode A or mode B, respectively}. The results of $\Delta\Phi$, $\Delta\mu$ and $W_{50}$ are listed in Table \ref{tab:PhnW} and plotted in Fig. \ref{fig:wph}.
\begin{deluxetable}{rccccl}[htbp]
\tablecaption{Phase offset and FWHM of the two modes \label{tab:PhnW}}
\tablecolumns{6}
\tablenum{3}
\tablewidth{0pt}
\tablehead{
 \colhead{$\nu$} & \colhead{$\Delta\Phi$} & \colhead{$\Delta\mu$} & \colhead{$W_{\rm 50,A}$} & \colhead{$W_{\rm 50,B}$} &\colhead{Ref.} \\
\colhead{(MHz)}  & \colhead{(deg.)} & \colhead{(deg.)} & \colhead{(deg.)} & \colhead{(deg.)} & \colhead{}
}
\startdata
 686    &	2.81(16)& 1.44(14)  & {6.46(15)}   & 6.24(15) & 2005a \\
1369    &	--      & 1.94(13)  & 7.34(12)   & 6.24(13) & 2005b \\
3100    &	--      & 1.86(17)  & 7.12(14)   & 6.06(15) & 2005c \\
 430    &	2.15(16)&   --      & 5.15(16)   & 4.73(16) & \citetalias{1980ApJ...239..310F} \\
 327    &	1.59(54)&   --      & 6.23(54)   & 7.26(54) & \citetalias{2014MNRAS.439.3951S} \\
1400    &	2.27(54)&   --      & 7.66(54)   & 6.33(54) & \citetalias{2014MNRAS.439.3951S} \\
 327    &	1.52(32)&   --      & 5.80(32)   & 6.84(32) & \citetalias{2016MNRAS.462.2518R} \\
 820    &	1.63(16)&   --      & 6.98(16)   & 6.30(16) & \citetalias{2016MNRAS.462.2518R} \\
\enddata
\tablecomments{The errors in the brackets were estimated by considering both the sampling interval and the dispersion smearing due to the finite channel bandwidth. $\Delta\Phi$ and $\Delta\mu$ are the phase offsets between the peaks of the original pulse profiles and of the best-fit Gaussians.}
\end{deluxetable}

The left panel of Fig.~\ref{fig:wph} \edit1 {presents scattering of the phase offset versus frequency. At 686 MHz, as the $\Delta\Phi$ estimate is almost double the $\Delta\mu$ estimate, and considering the profile skewness and relatively low signal-to-noise ratio of the profiles of the two modes, we cannot exclude the possibility that $\Delta\Phi$ is an overestimation and that the $\Delta\mu$ is an underestimation of the true phase offset. Therefore, we use an averaged value ($2.13\pm0.21$) of them, and plot it in the panel. The plot} does not show a regular frequency dependence for the phase offset. The offsets are scattering within a range from 1.52 to 2.27, with a weighted mean value of $1.91\pm0.07$. The averaged value ($2.13\pm0.21$) of the $\Delta\Phi$ and $\Delta\mu$ for 686 MHz was used in the above calculation, and the $\Delta\mu$ value may be an underestimation, due to the profile skewness. It suggests that the separation between the beam centers of the two modes is generally stable in the frequency range between 327 and 3100 MHz.
The right panel of Fig. \ref{fig:wph} shows an intriguing phenomenon that modes A and B have the opposite trends in the frequency dependence of pulse width, i.e. the pulse width of mode A generally increases with frequency, while that of mode B generally decreases with frequency. The difference mainly manifests in the low-frequency range between $\sim$300 and $\sim$700 MHz. This difference suggests that the broadening of pulse width of mode B at low frequencies, e.g. 327 MHz, is not mainly caused by interstellar scattering, but should be attributed to the intrinsic broadening.

A well-known empirical relationship, $W_{50}=A\nu^\alpha+B$ \citep{1991ApJ...377..263T,2002ApJ...577..322M,2014ApJS..215...11C}, called the Thorsett relationship, was employed to fit the data, except for the two data points at 430 MHz, because the published profiles in the earliest phase and the latest phase are not purely of modes A and B (\citetalias{1980ApJ...239..310F}). The best-fit relationships of modes A and B are  \edit1{$W_{50,\rm A} = -0.4(3)\nu^{-1.3(5)}+7.4(2)$} and \edit1{$W_{50, \rm B} = 0.1(1)\nu^{-2.0(1.5)}+6.1(1)$} deg, respectively. The two curves intersect at a frequency of about 500 MHz, meaning that the profile of mode A becomes wider than that of mode B above this frequency. Interestingly, combining the result here with that in subsection \ref{subsec:Spec}, both the inversions of the relative brightness and the relative pulse broadness between the two modes occur at very similar transition frequencies $\sim500$ MHz. However, noting that the FWHM of the profile in later phases is smaller than that of the profile in earlier phases at 430 MHz, we are aware that a small uncertainty may exist in the transition frequency.

\subsection{Correlation between the peak amplitude and phase for mode B} \label{subsec:Corr}

From the Gaussian fit to the sub-integrated profiles of modes A and B, we obtained the best-fit parameters for each profile, i.e. the amplitude $A$ and the phase $\mu$ of the profile peak and pulse width $\sigma$. The FWHM is then calculated by the equation \edit1{FWHM=$2\sqrt{2\ln 2}\sigma$}.

\begin{figure*}[htbp]
  \plotone{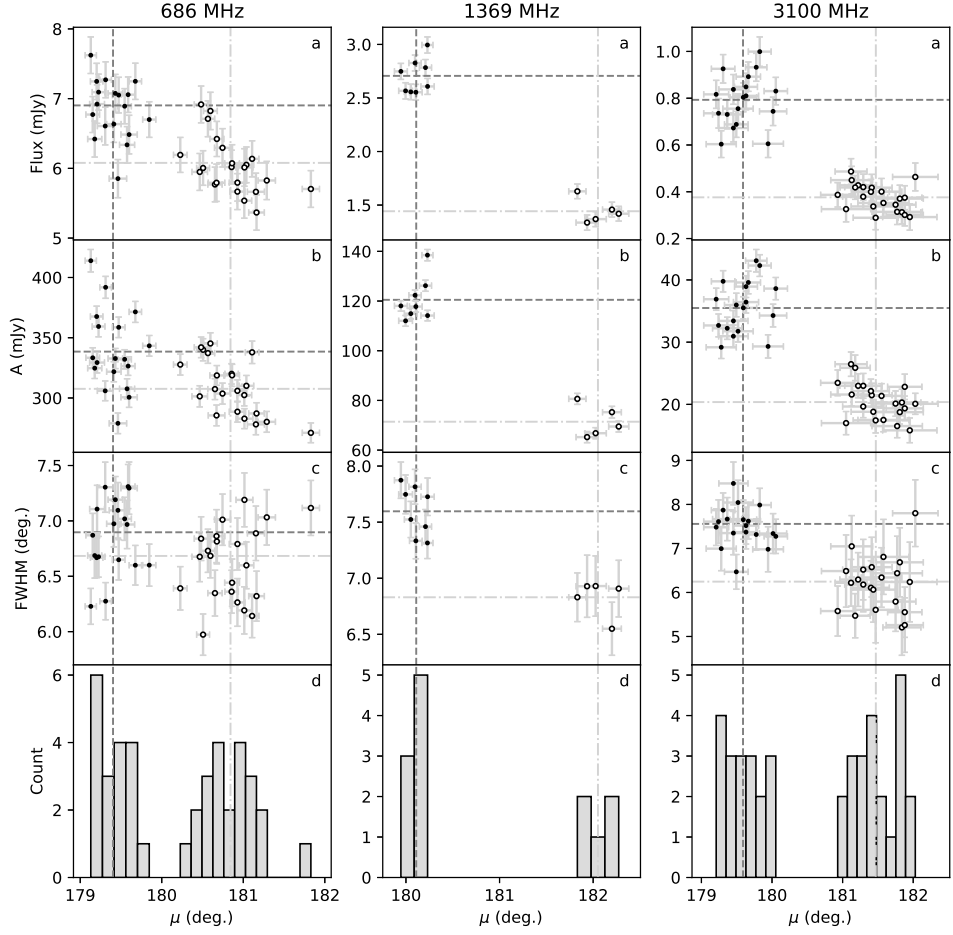}
   \caption{Distributions of (a) the flux, (b) the peak amplitude, and (c) the FWHM of the best-fit Gaussians for the subintegration profiles of modes A and B. The histograms of the phase of pulse peak ($\mu$) are shown in panels (d) for the three frequencies of the Parkes observations; see the text for details. A clear anticorrelation between the flux (or the peak amplitude) and the peak phase $\mu$ can be found for mode B at 686 and 3100 MHz. There is no such a clear anticorrelation for mode A.
   \label{fig:corr}}
\end{figure*}

The distributions of $\mu$ are shown in panels (d) of Fig. \ref{fig:corr} for the three frequencies, in which the two modes are clearly separated into two groups, with mode A in earlier phases and mode B in later phases. Two vertical dashed lines were plotted to represent the average peak phases for the two modes.

 Panels \edit1{(b)} of Fig. \ref{fig:corr} display the peak amplitude versus peak phase. Mode A is quite scattered, but mode B clearly shows an anticorrelation at 686 and 3100 MHz, which indicates that subintegration pulses are weaker when occurring at later phases. This correlation is also visible in the figures of time series of $A$ and $\mu$ in \citetalias{2014MNRAS.439.3951S} at 327 and 1400 MHz, although the authors did not conduct a quantitative analysis on this. Unfortunately, lacking enough subintegrations, no correlation can be confirmed here at 1.4 GHz.

Despite the scattered distribution of the FWHM (see panels \edit1{(c)}), the flux density of subintegration also follows an anticorrelation with $\mu$ at 686 MHz and 3100 MHz for mode B, as shown by panels (a). The simultaneous observation at these two frequencies enables us to \edit1{examine} whether the spectrum varies with the pulse phase that mode B occurs.

\section{Implication to spectral properties of the emission beams}
\label{sect:Impl}

The above differences in the frequency-dependent behaviors of intensity and pulse width probably imply a significant difference in the distributions of spectral index across the emission beams of the two modes. A direct way to probe this is to measure the spectral index across the pulse phase. However, such a study requires observations on the phase-aligned multifrequency profiles, which have not been carried out before\footnote{\citetalias{2016MNRAS.462.2518R} did show phase-aligned pulse tracks in Fig. 5 for two simultaneous observations at 327/150 MHz and 820/150 MHz, separately, but 327 and 820 MHz are not observed simultaneously, and in Fig. 6 the profiles at 327 and 820 MHz are not phase aligned.}. Therefore, in this paper, we tried to infer the spectral properties of the beam by the behaviors of intensity, pulse width, and phase offset.

\begin{figure*}[htbp]
  \plotone{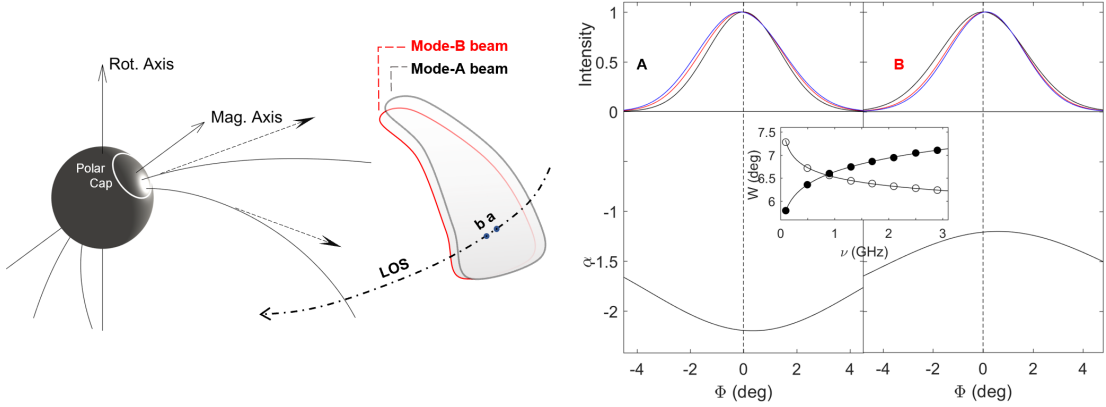}
   \caption{Left: schematic diagram of the emission beam of the two modes based on the fan-beam model. Right: simulated spectral-index ($\alpha$) variations across the pulse phase and the resultant frequency evolution of pulse profile at 300 MHz (black), 1.5 GHz (red), and 3.0~GHz (blue). The pulse profile and the shape of the spectral-index variation are both assumed to be Gaussian. The resultant frequency dependences of pulse width were plotted in the inset for modes A (solid circles) and B (open circles). \label{fig:beam} }
\end{figure*}

For the sake of convenience, we first plot a schematic diagram of the emission beams of two modes on the basis of the fan-beam model (\citealt{2010MNRAS.405..509D}; \citealt{2014ApJ...789...73W}). Assuming that the relativistic charged particles flowing out along a magnetic flux tube may generate wide-band radio emission, \citet{2014ApJ...789...73W} demonstrated that the resultant beam could be elongated in latitude, as shown schematically by the beam shape and the corresponding flux tube rooted on the light area in the polar cap in the left panel of Fig. \ref{fig:beam}. The two shaded beams represent mode A (in black) and mode B (in red), respectively, with the mode A beam being swept by the line of sight (LOS) earlier than the mode B beam. Here, the boundary of a shaded beam does not mean a cutoff of the emission, but represents a certain intensity level relative to the peak intensity in the beam, e.g. 5\%.

The sweeping of the LOS produces multifrequency pulse profiles. To illustrate how the frequency dependence of pulse profiles relates to the spectral variation in the emission beam (along the direction of LOS), a simulation of spectral-index distributions and the resultant pulse profiles at 300 MHz (black), 1.5 GHz (red), and 3 GHz (blue) are presented in the right panel of Fig. \ref{fig:beam}, with sub panels (A) and (B) for modes A and B, respectively. The variation of FWHM versus frequency is plotted in the inset, with the solid dots for mode A and the open circles for mode B. It is shown that the increasing trend of frequency dependence of the pulse width can arise from a type of inner-steep-outer-flat spectrum; namely, the spectrum becomes steeper from the edge to the center. On the contrary, a type of inner-flat-outer-steep spectrum can result in the decreasing trend of frequency dependence. This is consistent with a similar simulation carried out by \citet{2014ApJS..215...11C} to explain two opposite types of frequency dependence of pulse width (see Figure 11 of that paper). As the slight variation in the phase offset between the pulse peaks of the two modes probably suggest that the peak phases of the multifrequency profiles may be slightly different for either of the mode, we have set the flattest and steepest spectra to be slightly offset from the central phase ($\Phi=0$). The resultant shift of the peak can be seen from the simulated profiles at different frequencies, but this does not change the general frequency dependence of the pulse width, as long as the spectral index follows the forms displayed in the plot.

To summarize, the differences in the frequency-dependent behaviors of intensity and pulse width are very likely to be caused by two opposite types of spectral-index variation across the beam (in the longitudinal direction) for modes A and B.

\section{Conclusion and Discussion} \label{sect:Disc}

It is known that the emission of the young pulsar PSR J0614+2229 switches between two modes that are offset in pulse phase and differ in intensity. In this paper, the frequency dependence of emission properties, i.e. the intensity, pulse width, and phase offset are studied in details for the two modes by using multifrequency data that are both observed with the Parkes radio telescope and collected from the literature. Within a wide frequency range from 327 to 3100 MHz, the phase offset between the profile peaks of the two modes is roughly stable at most frequencies, with an average value of $1.95\pm0\fdg08$. Our analysis offers the following findings on the difference between the two modes.

(1) The mode occurring earlier in pulse phase (mode A) has a flatter spectrum than that occurring later in phase (mode B), with a difference in the spectral indices of $\sim0.5$. Mode A is weaker than mode B at low frequencies, but due to the spectral difference, it surpasses mode B at $\sim$500 MHz.

(2) Modes A and B follow different types of frequency dependence of pulse width. The FWHM of mode A increases with frequency, while that of mode B decreases with frequency. Our simulation suggests that two opposite types of spectral variation across the beam along the LOS can account for the difference in the frequency dependence of pulse width; namely, the spectrum becomes steeper from the edge to the center for mode A, while the opposite is true for mode B.

(3) For mode B, the peak amplitude and the flux density of subintegrated profile are anticorrelated with the phase of the pulse peak, indicating that the emission at earlier phases is relatively stronger. This anticorrelation is not seen in mode A.

The bursting emission at 327 MHz was discovered by \citetalias{2014MNRAS.439.3951S}, which is stronger than the other mode and show systematically decaying structure in intensity. It was once puzzling that the bursting emission occurs earlier in pulse phase at 820 MHz, but later in phase at 327 MHz \citepalias{2016MNRAS.462.2518R}. In fact, as found in this paper, the relatively stable phase offset between the two modes can exclude the possibility that the bursting emission translates from the trailing to the leading part of the beam as the frequency increases. Therefore, the change of relative brightness between the two modes can be well explained by the difference in their spectral indices. As for the frequency dependence of the decay behavior of the bursts, i.e. less noticeable for mode B but more noticeable for mode A at higher frequencies, it may be related to the evolution of the spectral index during the bursting period. \edit1{In this paper, there is a sign that sometimes, e.g. in the 46th subintegration at 686 MHz, mode A is brighter immediately after transition from mode B (see Fig. \ref{fig:raw}), which is in agreement to the bursting behavior at 327 MHz. However, the 1-min integration time of subintegrations inevitably leads to mixture of the bursting of mode A and the weak emission of mode B. This has prevented us to evaluate the spectral property of the bursting emission just after transition.} Future simultaneous observations at broad radio bands will help to understand this.

\acknowledgments

We appreciate the anonymous referee for a thorough and valuable comments. We thank Richard N. Manchester, Hao Tong, Shi Dai, Jumei Yao, Wenjun Huang, and Qihui Chen for helpful discussions. This work is supported by National Natural Science Foundation of China (Nos. 11178001, 11573008, U1631106). \edit1{Yanrong Zhang appreciates Guangzhou University--University of Padova Joint PhD Program}. The Parkes radio telescope is part of the Australia Telescope National Facility which is funded by the Australian Government for operation as a National Facility managed by CSIRO. This paper includes archived data obtained through the Australia Telescope Online Archive and the CSIRO Data Access Portal (\url{http://data.csiro.au}).

%

%
%

\bibliography{PSRJ0614+2229}

\begin{thebibliography}{}
\expandafter\ifx\csname natexlab\endcsname\relax\def\natexlab#1{#1}\fi
\providecommand{\url}[1]{\href{#1}{#1}}
\providecommand{\dodoi}[1]{doi:~\href{http://doi.org/#1}{\nolinkurl{#1}}}
\providecommand{\doeprint}[1]{\href{http://ascl.net/#1}{\nolinkurl{http://ascl.net/#1}}}
\providecommand{\doarXiv}[1]{\href{https://arxiv.org/abs/#1}{\nolinkurl{https://arxiv.org/abs/#1}}}

\bibitem[{{Chen} \& {Wang}(2014)}]{2014ApJS..215...11C}
{Chen}, J.~L., \& {Wang}, H.~G. 2014, \apjs, 215, 11,
  \dodoi{10.1088/0067-0049/215/1/11}

\bibitem[{{Dai} {et~al.}(2017){Dai}, {Johnston}, \&
  {Hobbs}}]{2017MNRAS.472.1458D}
{Dai}, S., {Johnston}, S., \& {Hobbs}, G. 2017, \mnras, 472, 1458,
  \dodoi{10.1093/mnras/stx2033}

\bibitem[{{Davies} {et~al.}(1972){Davies}, {Lyne}, \&
  {Seiradakis}}]{1972Natur.240..229D}
{Davies}, J.~G., {Lyne}, A.~G., \& {Seiradakis}, J.~H. 1972, \nat, 240, 229,
  \dodoi{10.1038/240229a0}

\bibitem[{{Dyks} {et~al.}(2010){Dyks}, {Wright}, \&
  {Demorest}}]{2010MNRAS.405..509D}
{Dyks}, J., {Wright}, G.~A.~E., \& {Demorest}, P. 2010, \mnras, 405, 509,
  \dodoi{10.1111/j.1365-2966.2010.16462.x}

\bibitem[{{Ferguson} \& {Boriakoff}(1980)}]{1980ApJ...239..310F}
{Ferguson}, D.~C., \& {Boriakoff}, V. 1980, \apj, 239, 310,
  \dodoi{10.1086/158112}

\bibitem[{{Geyer} {et~al.}(2017){Geyer}, {Karastergiou}, {Kondratiev},
  {Zagkouris}, {Kramer}, {Stappers}, {Grie{\ss}meier}, {Hessels}, {Michilli},
  {Pilia}, \& {Sobey}}]{2017MNRAS.470.2659G}
{Geyer}, M., {Karastergiou}, A., {Kondratiev}, V.~I., {et~al.} 2017, \mnras,
  470, 2659, \dodoi{10.1093/mnras/stx1151}

\bibitem[{{Gupta} {et~al.}(1994){Gupta}, {Rickett}, \&
  {Lyne}}]{1994MNRAS.269.1035G}
{Gupta}, Y., {Rickett}, B.~J., \& {Lyne}, A.~G. 1994, \mnras, 269, 1035,
  \dodoi{10.1093/mnras/269.4.1035}

\bibitem[{{Helfand} {et~al.}(1980){Helfand}, {Taylor}, {Backus}, \&
  {Cordes}}]{1980ApJ...237..206H}
{Helfand}, D.~J., {Taylor}, J.~H., {Backus}, P.~R., \& {Cordes}, J.~M. 1980,
  \apj, 237, 206, \dodoi{10.1086/157860}

\bibitem[{{Hobbs} {et~al.}(2011){Hobbs}, {Miller}, {Manchester}, {Dempsey},
  {Chapman}, {Khoo}, {Applegate}, {Bailes}, {Bhat}, {Bridle}, {Borg}, {Brown},
  {Burnett}, {Camilo}, {Cattalini}, {Chaudhary}, {Chen}, {D'Amico},
  {Kedziora-Chudczer}, {Cornwell}, {George}, {Hampson}, {Hepburn}, {Jameson},
  {Keith}, {Kelly}, {Kosmynin}, {Lenc}, {Lorimer}, {Love}, {Lyne}, {McIntyre},
  {Morrissey}, {Pienaar}, {Reynolds}, {Ryder}, {Sarkissian}, {Stevenson},
  {Treloar}, {van Straten}, {Whiting}, \& {Wilson}}]{2011PASA...28..202H}
{Hobbs}, G., {Miller}, D., {Manchester}, R.~N., {et~al.} 2011, \pasa, 28, 202,
  \dodoi{10.1071/AS11016}

\bibitem[{{Hotan} {et~al.}(2004){Hotan}, {van Straten}, \&
  {Manchester}}]{2004PASA...21..302H}
{Hotan}, A.~W., {van Straten}, W., \& {Manchester}, R.~N. 2004, \pasa, 21, 302,
  \dodoi{10.1071/AS04022}

\bibitem[{{Jankowski} {et~al.}(2018){Jankowski}, {van Straten}, {Keane},
  {Bailes}, {Barr}, {Johnston}, \& {Kerr}}]{2018MNRAS.473.4436J}
{Jankowski}, F., {van Straten}, W., {Keane}, E.~F., {et~al.} 2018, \mnras, 473,
  4436, \dodoi{10.1093/mnras/stx2476}

\bibitem[{{Lewandowski} {et~al.}(2011){Lewandowski}, {Kijak}, {Gupta}, \&
  {Krzeszowski}}]{2011A&A...534A..66L}
{Lewandowski}, W., {Kijak}, J., {Gupta}, Y., \& {Krzeszowski}, K. 2011, \aap,
  534, A66, \dodoi{10.1051/0004-6361/201116850}

\bibitem[{{Mitra} \& {Rankin}(2002)}]{2002ApJ...577..322M}
{Mitra}, D., \& {Rankin}, J.~M. 2002, \apj, 577, 322, \dodoi{10.1086/342136}

\bibitem[{Nowakowski(1992)}]{nowakowski_1992}
Nowakowski, L.~A. 1992, International Astronomical Union Colloquium, 128, 280,
  \dodoi{10.1017/S0002731600155337}

\bibitem[{{Rajwade} {et~al.}(2016){Rajwade}, {Seymour}, {Lorimer},
  {Karastergiou}, {Serylak}, {McLaughlin}, \&
  {Griessmeier}}]{2016MNRAS.462.2518R}
{Rajwade}, K., {Seymour}, A., {Lorimer}, D.~R., {et~al.} 2016, \mnras, 462,
  2518, \dodoi{10.1093/mnras/stw1858}

\bibitem[{{Seymour} {et~al.}(2014){Seymour}, {Lorimer}, \&
  {Ridley}}]{2014MNRAS.439.3951S}
{Seymour}, A.~D., {Lorimer}, D.~R., \& {Ridley}, J.~P. 2014, \mnras, 439, 3951,
  \dodoi{10.1093/mnras/stu250}

\bibitem[{{Thorsett}(1991)}]{1991ApJ...377..263T}
{Thorsett}, S.~E. 1991, \apj, 377, 263, \dodoi{10.1086/170355}

\bibitem[{{van Straten} {et~al.}(2012){van Straten}, {Demorest}, \&
  {Oslowski}}]{2012AR&T....9..237V}
{van Straten}, W., {Demorest}, P., \& {Oslowski}, S. 2012, Astronomical
  Research and Technology, 9, 237.
\newblock \doarXiv{1205.6276}

\bibitem[{{Wang} {et~al.}(2014){Wang}, {Pi}, {Zheng}, {Deng}, {Wen}, {Ye},
  {Guan}, {Liu}, \& {Xu}}]{2014ApJ...789...73W}
{Wang}, H.~G., {Pi}, F.~P., {Zheng}, X.~P., {et~al.} 2014, \apj, 789, 73,
  \dodoi{10.1088/0004-637X/789/1/73}

\bibitem[{{Weisberg} {et~al.}(1999){Weisberg}, {Cordes}, {Lundgren}, {Dawson},
  {Despotes}, {Morgan}, {Weitz}, {Zink}, \& {Backer}}]{1999ApJS..121..171W}
{Weisberg}, J.~M., {Cordes}, J.~M., {Lundgren}, S.~C., {et~al.} 1999, \apjs,
  121, 171, \dodoi{10.1086/313189}

\end{thebibliography}


\end{document}